\documentclass[11pt]{article}
\usepackage{pst-all}
\usepackage{pstricks}
\usepackage{pstcol,pst-fill,pst-grad}
\usepackage{amsfonts}
\usepackage{graphicx}
\usepackage{amsmath}
\usepackage[normalem]{ulem}

\usepackage{tikz}
\usetikzlibrary{matrix}
\tikzset{squarestyle/.style={align=center,draw, text centered,minimum width=1.5cm, minimum height=1cm}}
\tikzset{matrixstyle/.style={matrix of nodes,nodes in empty cells,ampersand replacement=\&,column sep=0.5em,row sep=1ex}}

\makeatletter \@addtoreset{equation}{section}

\newcommand{\be}{\begin{equation}}
\newcommand{\ee}{\end{equation}}
\newcommand{\bea}{\begin{eqnarray}}
\newcommand{\eea}{\end{eqnarray}}
\begin{document}
\date{}
\title{
\textbf{    Qubit and  Fermionic Fock Spaces  from Type II Superstring Black Holes  }\\
\textbf{   } }
\author{ A.  Belhaj$^{1}$,  M. Bensed$^{2}$, Z. Benslimane$^{2}$,  M. B.
Sedra$^{2,3}$, A. Segui$^{4}$
\hspace*{-8pt} \\
\\
{\small $^{1}$D\'epartement de Physique, LISRT,  Facult\'e
Polydisciplinaire, Universit\'e Sultan Moulay Slimane}\\{ \small
B\'eni Mellal, Morocco }
\\ {\small $^{2}$    D\'{e}partement de Physique, LabSIMO,  Facult\'{e}
des Sciences, Universit\'{e} Ibn Tofail }\\{ \small K\'{e}nitra,
Morocco}
\\ {\small $^{3}$
\'Ecole Nationale des Sciences Appliqu\'ees, Universit\'{e} Ibn
Tofail,  K\'{e}nitra, Morocco}
\\ {\small $^{4}$   Departamento de F\'isica Te\'orica, Universidad de
Zaragoza, E-50 009-Zaragoza, Spain}  }  \maketitle

\begin{abstract}
Using Hodge diagram combinatorial data, we  study  qubit and
fermionic Fock spaces  from  the point of view of type II superstring black  holes based on
complex compactifications.
  Concretely, we establish a
one-to-one correspondence between qubits, fermionic  spaces and
extremal   black holes  in maximally supersymmetric supergravity
obtained from  type II superstring on complex toroidal and
Calabi-Yau compactifications. We interpret the
differential  forms of   the $n$-dimensional complex  toroidal
compactification
 as states  of  $n$-qubits
encoding  information on  extremal black hole charges. We show that
there are  $2^n$ copies of $n$ qubit systems which  can be split as
$ 2^n=2^{n-1}+2^{n-1}$. More precisely, $2^{n-1}$ copies are
associated  with even D-brane charges in type IIA superstring and
the other $2^{n-1}$ ones  correspond to   odd D-brane charges  in
IIB superstring. This  correspondence is generalized to a class of
Calabi-Yau manifolds. In  connection with  black hole charges in
type IIA superstring,  an $n$-qubit system has  been obtained from a
canonical line bundle of $ n$ factors of one dimensional projective
space $ \mathbb{CP}^1.$
\end{abstract}


\thispagestyle{empty}

\newpage \setcounter{page}{1} \newpage

\section{Introduction}

Extremal black  holes   have  been  extensively investigated  in the
framework of string theory and related topics including M and
F-theories  compactified on Calabi-Yau manifolds \cite{1,2,3,4}.
These black solutions have been studied using  attractor mechanism
and topological string theory \cite{5,6,7,8,9}.
 In attractor
mechanism scenario, the scalar fields can be fixed in terms of the
black  hole charges by extremising the
 corresponding potential with respect to stringy  moduli  obtained from the
compactification of higher dimensional theories.  Moreover, the
corresponding entropy functions have been  calculated  using the
U-duality symmetry acting on the invariant black  hole charges of
the compactified theories. In this issue, the Calabi-Yau
compactifications have been investigated   producing various results
dealing with  black  holes in type II superstrings  using D-brane
physics \cite{10,11}.

Besides these activities, black  holes in string theory
compactification  have been connected with quantum information using
the qubit formalism [12-22]. Concretely, a possible link between the
$N=2$ STU black hole obtained from type II superstring and three
qubits has been given in \cite{15,16}. The main recent works, in
quantum information theory, are trying to connect quantum
entanglement and invariant theory, by  classifying orbits of entangled
states in multipartite configurations of qubit systems,  using
multidimensional matrix invariants called hyperdeterminants,  which were found by
Cayley as extensions for the usual determinants of 2-dimensional
matrices of classical algebra.  In fact, the notion of hyperdeterminant
 has been relevant in theso called black hole/qubit correspondence.  \\
Some of these
connections have been extended to construct superqubits using the
compactification on supermanifolds \cite{19,23}. It has also been
noted a relationship between qubits  and Segre varieties. In a
related work, the Segre embedding has been used to give the
connection between geometry and concurrence which describes
entanglement of formation for bipartite pure states qubit systems
\cite{20}. More recently, qubit systems have been embedded into
fermionic Fock space \cite{24} by using the classification of pure
states entanglement in the framework of spinors classification, by which
a physical interpretation has been  found. \\

The main goal of  this  work  is to contribute to these activities
by investigating  qubit and  fermionic Fock spaces  from   type II
superstring black  hole framework. Using combinatorial data provided
by Hodge diagrams, we establish a one-to-one correspondence between
qubits, fermionic spaces and extremal   black holes  in maximally
supersymmetric supergravity obtained from  type II superstring on
complex toroidal and Calabi-Yau compactifications. To this aim, we
interpret the differential  forms of   the $n$-dimensional complex
toroidal compactification as states  of $n$-qubits encoding
information on  extremal black hole charges.\\
 We claim that $2^n$
copies of $n$ qubit systems can be split as $ 2^n=2^{n-1}+2^{n-1}$.
In this decomposition, $2^{n-1}$ copies correspond to  even D-brane
charges in type IIA superstring
 and the other $2^{n-1}$ ones  are associated with   odd D-brane charges  in  IIB superstring.
This  toroidal compactification  correspondence  can be  generalized
to a class of Calabi-Yau manifolds. In particular, an $n$-qubit
system has been  associated with  a canonical line bundle of $ n$
factors of one dimensional projective space $\mathbb{CP}^1$  which encodes data
of  type IIA  black hole charges.

The organization of the paper is as follows. In section 2, we
reconsider the study of the extremal black  holes in type II
superstrings.  Section 3 concerns  the  correspondence between
qubits and   balck holes. In section 4, we  study  qubits and
fermionic Fock space from  type II superstrings on toroidal  complex
compactifications. The generalization to  a class of Calabi-Yau
manifolds is given in section 5.  The last section is devoted to
conclusions and open questions.

\section{ Black holes in type II superstrings  on complex
manifolds}
 In this section we discuss  black holes in type II
superstrings on complex compact manifolds. Before going
further, we recall
 that compact complex manifolds play an important role
in string theory. We will be concerned with the cohomology structure  properties
which enclose information on the corresponding black hole charges using
D-brane objects. \\
In the context of differential geometry, an $n$-dimensional
 compact complex manifolds $M^n$   involves complex and real  forms carrying a rich
 structure. These forms play a crucial role in the string theory
 compactification and  related physics, including the geometric construction of  gauge theories\cite{25},
 and they can be summarized in a nice graphic representation called Hodge
 diagram. The latter  is  a relevant  piece for the determination of lower
 dimensional spectrum of string theory and the associated black hole charges. It has been remarked that
 the studied manifolds are the K\"{a}hlerian ones  such as
 $T_\mathbb{C}^n=T^{2n}$, $\mathbb{CP}^n$ and the Calabi-Yau manifolds $CY^n$.   \\
In order to interpret qubits  and  fermionic Fock spaces  in terms
of  differential forms,  we   emphasize the geometric properties of
cohomology space. Without loss of generality, consider an
$n$-dimensional compact complex manifold $M^n$. The space of
$k$-forms $\Omega^{k}(M)^{\mathbb{C}}$ is decomposed as
\begin{equation}
\Omega^{k}(M)^{\mathbb{C}}=\bigoplus_{p+q=k} \Omega^{p,q}(M).
\end{equation}
It is recalled that any element   $\omega$  of $ \Omega^{p,q}(M)$ is
an antisymmetric tensor with $p$ holomorphic and $q$
anti-holomorphic components. It is recalled that $\omega$ of
$\Omega^{p,0}(M)$ satisfies $\bar{\partial}\omega=0$  where
$\bar{\partial}$ is a Dolbeault operator. A sequence of
$\mathbb{C}$-linear maps  reads as
\[
\Omega^{p,0}(M)\stackrel{\bar{\partial}}{\rightarrow} \quad \Omega^{p,1}(M)\stackrel{\bar{\partial}}
{\rightarrow} \quad ... \quad \stackrel{\bar{\partial}}{\rightarrow} \quad \Omega^{p,n-1}(M)\stackrel{\bar{\partial}}{\rightarrow} \quad \Omega^{p,n}(M)
\]
which is the Dolbeault complex. From the antisymmetry property of
differential forms, $\bar{\partial}$ verifies
 the nilpotency property $\bar{\partial}^2=0$ which leads to the construction
of  the $(p,q)^{\text{th}} \bar{\partial}$ cohomology group, that
forms also a complex vector space. It is defined as
\begin{equation}
H_{\bar{\partial}}^{p,q}(M)\equiv
\frac{Z_{\bar{\partial}}^{p,q}(M)\equiv ker
\bar{\partial}}{B_{\bar{\partial}}^{p,q} \equiv im \bar{\partial}}
\end{equation}
where $Z_{\bar{\partial}}^{p,q}(M)$ is the set of $(p,q)$-cocycles
on $M$, and $B_{\bar{\partial}}^{p,q}(M)$ is the set of
$(p,q)$-coboundaries. Later, we shall
 drop the $\bar{\partial}$ symbol, since we will be concerned only
by  the cohomology space of K\"{a}hlerian manifolds.

It is known that $H^{p,q}(M)$ measures the topological
non-triviality of complex manifolds, since its dimension is the
hodge number $h^{p,q}$, and we have $dim_{\mathbb{C}}~H^{p,q}(M)=
h^{p,q}$. An enlightening remark is that the cohomology space $H$
of a complex manifold has an $\mathbb{N}$-graded ring structure
which reads as
\begin{equation}
 H=\bigoplus_{k\in \mathbb{N}} H^k(M) \qquad \text{where} \qquad H^k(M)=
 \bigoplus_{k=p+q}H^{p,q}(M).
\end{equation}
The Hodge numbers summarize the information needed to build  the
qubit spaces associated with black hole charges in type II
superstrings. They spread in a two dimensional diagram commonly
called Hodge Diamond. Generally, it takes
    the following form
\[
\begin{array}{ccccc}
 &  & h^{0,0}\\
 & h^{1,0} & \vdots & h^{1,0}\\
h^{n,0} & ... &  & ... & h^{0,n}\\
 & h^{n,n-1} & \vdots & h^{n-1,n}\\
 &  & h^{n,n}
\end{array}
\]

Accordingly, the Hodge diamond for K\"{a}hler manifolds has
vertical and horizontal symmetries, and therefore has the following
properties:
\begin{itemize}
    \item $h^{p,q}=h^{q,p}$
    \item  $h^{p,q}= h^{n-p,n-q}$
    \item the number of independent Hodge numbers are $(\frac{1}{2}n+1)^2$ if $n$ even,
    and $\frac{1}{4}(n+1)(n+3)$ if $n$ is odd.
\end{itemize}
More constraints  can be implemented by  imposing other geometric
conditions including  the Calabi-Yau one.

To make  a possible contact with qubit systems, we consider a
special compact geometry
\begin{equation}
M^n=\underbrace{{\bf T}^2\times {\bf T}^2\times\ldots\times {\bf
T}^2\times {\bf T}^2}_n
\end{equation}
where  $ {\bf T}^2$  is 2-dimensional torus. It is convenient
to use the  complex coordinates. Indeed, $ {\bf T}^2$  is defined by
the following identifications and constraints
\begin{equation}
 z \equiv z + 1,\qquad  z \equiv z + \imath,\qquad  i^2=-1
 \end{equation}
and we will use the complex  notation $M^n=T^n_\mathbb{C}$.\\
The cohomology classes associated with  the holomorphic and the  anti
holomorphic $(p,q)$ forms are
\begin{equation}
1 ~,~dz_i~,~ \overline{dz}_j~,~ dz_i\wedge {dz}_j~,~dz_i\wedge
\overline{dz}_j~,~\ldots~, ~ dz_1\wedge \ldots dz_n \wedge
\overline{dz}_1\wedge \ldots \overline{dz}_n
\end{equation}
where $i,j,\ldots = 1,\ldots,n$. The corresponding $(n+1)^2$  Hodge  numbers can be listed in the
previous Hodge diagram. A close inspection shows that the total number
of forms on such  a manifold is
\begin{equation}
\sum_{k=0,1,\ldots,n}\sum_{k=p+q}{h^{p,q}}= 2^{2n}=2^{n}\times
2^{n}.
\end{equation}
There are many ways to interpret this number. In connection with
quantum information, this  indicates  that there are  $2^n$ copies
of $n$ qubit systems.  It has been remarked that these systems  can
be linked with  black objects  which  can be obtained from type II
superstring D-branes wrapping  non trivial cycles of
$T^n_\mathbb{C}$. It has been shown that the near horizon
of these black objects can be given by
the product of Anti-de-Sitter spaces and  spheres
\begin{equation}
Ads_{p+2}\times S^{6-2n-p},
\end{equation}
 where integers  $n$ and $p$  should  satisfy the following constraint
\begin{equation}
2 \leq 6-2n-p.
\end{equation} The electric/magnetic duality  requires that the $p$-dimensional electric black   branes and the $q$-dimensional
magnetic black branes are related by
\begin{equation}
\label{gd} p+q=6-2n.
\end{equation}
This  condition  can  generate  many black  solutions  which can be
classified  by the values  of $(p,q)$. However, we discuss here only
the case associated with $(p,q)=(0,6-2n)$ describing  electric
charged black holes. Indeed,  the compactification  on $T^n_\mathbb{C}$  may
produce the black  holes  configurations in $10-2n$ dimensional
maximally supersymmetric supergravity coupled to abelian gauge
symmetries associated with the { \bf NS-NS} and {\bf R-R} bosonic
fields of various ranks in type II superstrings \cite{25}.
Concretely, the black holes can be constructed using  the following
type II D-brane configurations
\begin{eqnarray*}
\mbox{Type IIA}&:& \mbox{D0-branes}, \quad \mbox{D2-branes}, \quad
\mbox{D4-branes},  \quad  \mbox{D6-branes},  \quad
\mbox{D8-branes}\\ \mbox{Type IIB}&:& \mbox{D1-branes}, \quad
\mbox{D3-branes}, \quad \mbox{D5-branes},  \quad  \mbox{D7-branes},
\quad \mbox{D9-branes}.
\end{eqnarray*}
Connections have been made recently between entropy of black holes,
in string theory, and qubit entanglement in quantum information
theory. These connections go under the name of the Black Hole Qubit
Correspondence (BHQC) \cite{15,16} which will be discussed in the
following sections

\section{Black Hole/Qubit Correspondence}
It is recalled that the qubit is a  building block,  in quantum
 information  theory,  which  has been extensively investigated using
  different physical and  mathematical approaches \cite{26,27,28}.
  It is a two level system which can be
 associated, for instance, with the electron  in the hydrogen atom.
 The general state of  a single  qubit is usually given by the
 Dirac notation as follows
\begin{equation}
|\psi\rangle=a_0|0\rangle+a_1 |1\rangle
\end{equation}
 where $a_i$  are complex  coefficients  satisfying the normalization
condition
\begin{equation}
|a_0|^2+|a_1 |^2=1.
\end{equation}
  This equation can be  interpreted  geometrically in terms of the  so called Bloch
sphere. The two qubits are four states systems. In this case,
the most general state  reads
\begin{equation}
|\psi\rangle=a_{00}|00\rangle+a_{10}
|10\rangle+a_{01}|01\rangle+a_{11} |11\rangle
\end{equation}
where $a_{ij}$  are complex numbers satisfying the normalization
condition
\begin{equation}
|a_{00}|^2+|a_{10}|^2+|a_{01}|^2+|a_{11}|^2=1,
\end{equation}
describing a   three dimensional complex projective space $\mathbb{CP}^3$
generalizing the Bloch sphere. This  analysis can be extended  to
$n$-qubits associated with  $2^n$ configuration states using the
same binary notation.  It is observed  that the  3-qubit systems can
be   represented by STU  black hole  charges in type IIB
superstring on $T^6$ \cite{15,16}. The connection has been
established using the entropy  formulae.  Indeed,  it is recalled
that the entropy can be defined in two different physical contexts.\\
In statistical mechanics, entropy is the measure of the number
 of available quantum states. This interpretation is well
motivated by string theory. In a  thermodynamical system, entropy is
associated with  the measure of order or disorder, which can only
increase according to the second law. Hawking showed that one can
assign this quantity to a black hole, which can radiate energy
 due to quantum mechanical effects \cite{29}. The entropy is then linked
to the event horizon area by Bekenstein-Hawking formulae
\begin{equation}
S_{BH}= \frac{c^3A}{4G\hbar} \propto \frac{1}{4}A.
\end{equation}
This shows a first interplay between a thermodynamic quantity of
quantum mechanical origin and a geometric quantity from the
classical theory of gravity. Moreover, it has been  revealed  that
the entropy of a black hole is related to the entanglement measure
of qubits  using Cayley's hyperdeterminant\cite{30}. To see that,
one considers the case of a pair of qubits $\{A,B\}$ described by a
general state
\begin{equation}
|\Psi \rangle = a_{AB}|AB\rangle \qquad   A,B = 0, 1
\end{equation}
where the Enstein summation convention is used. Then,  the bipartite
entanglement of $A$  and $B$ is given  in terms of  the 2-tangle
$\tau_{AB}$
\begin{equation}
\tau_{AB}= 4|deta_{AB}|^2=4|a_{00}a_{11}-a_{01}a_{10}|^2 = 4 |det
\rho_A|=4|det \rho_B|
\end{equation}
where $\rho_A$ and $\rho_B$ are respectively the partial trace of
$\rho_{AB}$ over $A$ and $B$.   The 2-tangle $\tau_{AB}$ is invariant under
$SL(2) \times SL(2)$
 stochastic local operations and classical communication(SLOCC). It
 is also invariant  under the  permutations of $A$ and $B$. \\
To discuss the case of  3-qubit system \{A,B,C\} described by a
general state $|\Psi \rangle= a_{ABC}|ABC\rangle$, we need to
introduce  Cayley's hyperdeterminant  considered as  a
generalization of a determinant from matrices to multidimensional
hypermatrices. The simplest version of hypermatrix A  of format $2
\times 2 \times 2$ can be visualized as a matrix  for which the 8
entries are placed in the corner of a cube. The hypermatrix
$\mbox{A}_{2\times2\times2}\equiv\left[a_{ABC}\right]$ where  $A,B,C
= 0,1$, can be represented in two dimensions by frontal slices of
the cube (where $k$ encodes the two slices)
\[
\mbox{A}=\left[\begin{array}{cc}
a_{000} & a_{010}\\
a_{100} & a_{110}
\end{array}\mid\begin{array}{cc}
a_{001} & a_{011}\\
a_{101} & a_{111}
\end{array}\right]
\]
Cayley's hyperdeterminant denoted $\mbox{Det A}$ (with
capital ``D'' to make distinction with usual determinant) is defined
as follows
\begin{equation}
\text{Det A} \equiv-\frac{1}{2}\epsilon^{A_{1}A_{3}}\epsilon^{A_{2}A_{4}}
\epsilon^{B_{1}B_{2}}\epsilon^{B_{3}B_{4}}\epsilon^{C_{1}C_{2}}\epsilon^{C_{3}C_{4}}a_{A_{1}B_{1}C_{1}}a_{A_{2}B_{2}C_{2}}a_{A_{3}B_{3}C_{3}}a_{A_{4}B_{4}C_{4}}
\end{equation}
It is a homogeneous quartic polynomial composed by the entries
$a_{ABC}$, which acts as discriminant for trilinear form $A(X,Y,Z)$.
The condition of discriminant $\mbox{Det A=0}$ is equivalent to the
non-trivial vanishing of the trilinear form,  which gives the
singular points of the latter other than the origin.  $\mbox{Det A}$
can be written explicitly as
\begin{eqnarray}
\text{Det A} &=&
a_{000}^{2}a{}_{111}^{2}+a_{001}^{2}a_{110}^{2}+a_{010}^{2}a_{101}^{2}+a_{011}^{2}a_{100}^{2}
\nonumber\\
&&-2(a_{000}a_{001}a_{110}a_{111}+a_{000}a_{010}a_{101}a_{111}
+a_{000}a_{011}a_{100}a_{111} \nonumber\\
&&+a_{001}a_{010}a_{101}a_{110}
+a_{001}a_{011}a_{100}a_{110}+a_{001}a_{010}a_{100}a_{111})
\nonumber\\
&&+4(a_{000}a_{011}a_{101}a_{110}+a_{001}a_{010}a_{100}).
\end{eqnarray} This polynomial involves an interesting physical
interpretation in terms of  the STU   black hole embedded  in type
II superstrings. The corresponding theory involves 4 photons
producing
 8  charges. More precisely, the bosonic sector consists of gravity coupled
 to 4 photons and 3 complex scalar fields denoted  S, T and U playing a role
 of string dualities. The equations of motion carry the same symmetries occurring
in the definition of the hyperdeterminant. Namely, they display a discrete
triality that interchanges S, T and U. Moreover, they involve an $SL(2)_S \times SL(2)_T \times SL(2)_U$
 symmetry \cite{21}, and the solution of an STU black hole in the case of spherical symmetry
is given in terms of 8 charges $(q_0, q_1 , q_2, q_3, p^0, p^1, p^2, p^3)$.\\
 Considering an extremal STU black hole, with vanishing surface gravity and minimal mass
compatible with the given charges, the square of the entropy is
proportional to a quartic polynomial of $q_0, q_1 , q_2, q_3, p^0,
p^1, p^2, p^3$  \cite{15,16}
\begin{equation}
\begin{split}
S^2 = & \quad \pi^2 \{ -(p^0q_0 + p^1q_1 + p^2q_2+ p^3q_3)^2 \\ &+
4((p^1q_1)(p^2q_2)+(p^1q_1)(p^3q_3)+ (p^3q_3)(p^2q_2) +q_0p^1p^2p^3
- p^0q_1q_2q_3) \}.
\end{split}
\end{equation}
Under a suitable correspondence, the 8 charges of  the  STU black
hole are linked to the components $a_{ABC}$ of 3-qubit system
$\{A,B,C\}$ through the entropy formulae. By taking the following
identification
\begin{center}
$\left(\begin{array}{c}
q_{0}\\
q_{1}\\
q_{2}\\
q_{3}\\
p^{0}\\
p^{1}\\
p^{2}\\
p^{3}
\end{array}\right) \qquad \longleftrightarrow \qquad \left(\begin{array}{c}
a_{000}\\
-a_{001}\\
-a_{010}\\
-a_{100}\\
a_{111}\\
a_{110}\\
a_{101}\\
a_{011}
\end{array}\right),$
\end{center}
 the correspondence between the
3-tangle $\tau_{ABC}$ and the black hole entropy  can be established through
\begin{equation}
S = \pi \sqrt{|Det a_{ABC}|} = \frac{\pi}{2} \sqrt{\tau_{ABC}}.
\end{equation}
In type IIB superstring, the black hole charges   can be obtained
from D3-branes wrapping cycles on internal geometry. These charges
are obtained from  the reduction of  the real 5-form $F_5$, which is
the gauge invariant field strength associated with the $RR$  gauge field
4-form   coupled to D3-branes. The corresponding   gauge invariant
field strengths are
\begin{equation}
\mathcal{F}_{2}^{\alpha }=d{A}^{\alpha },\qquad  \alpha=0,1,2,3
\end{equation}%
The gauge fields ${A}_{\mu }^{\alpha}$  are obtained  from the
compactification of the gauge 4-form on 3-cycles. The integration of
the field strength $\mathcal{F}_{2}^{\alpha }$ and  its dual Hodge
form $\mathcal{F^\star}_{2}^{\alpha }$ throughout the sphere
 $S_{\infty }^{2}$  gives the  electric  $q^{\alpha }$  and magnetic
charges $p^{\alpha }$ which are  given respectively by
\begin{equation}
q^{\alpha }=\int_{S_{\infty }^{2}}{\mathcal{F^\star}_{2}}^{\alpha
},\qquad \qquad p^{\alpha }=\int_{S_{\infty
}^{2}}\mathcal{F}_{2}^{\alpha },\qquad
 \alpha=0,1,2,3.
\end{equation}
These  charges  can be also obtained  using T-duality, transforming
D3-branes to even D-branes in type IIA superstring. Geometrically,
it transforms 3-cycles to even cycles in  the internal geometry
encoding black hole charges. \\
The dictionary between  black   hole
charges and qubits  is still growing to establish a complete
picture. In this sense, we intend to propose additional elements to
enrich such a  dictionary.

\section{ Qubit and Fermionic spaces from  type II superstring compactifications }
In this section, we show that qubits and fermionic Fock spaces can
be embedded in type II superstrings on the compact complex manifold
$T^n_\mathbb{C}$ using black hole charges associated with
$(p,q)$-forms. In fact, the structure of the space $H$ can be
associated  with a vector space   on which rely  D-brane charges in
type II superstrings. This  can be explored  to approach qubits from
string Hodge diagrams.
 A close inspection shows that  this  structure
 can be associated with the  fermionic Fock space in which $2^n$ copies
 of $n$-qubit systems are embedded. Before giving the connection,  it is worth
 noting that  the  Fock space is a Hilbert space completion,
 on which all possible states of $n$-identical particles are
 represented. It  reads as
\begin{equation}
\mathcal{F}(\mathcal{H})={\bigoplus_{n \in \mathbb{N}}S_\epsilon
\mathcal{H}^{\otimes n}} \qquad \text{where} \qquad \epsilon=\pm
\end{equation}
 where $S_\epsilon$ corresponds to the symmetrizer tensor associated with
bosonic statistics if $\epsilon=+$, or to  the antisymmetrizer tensor associated
with fermions if $\epsilon=-$.\\
The fermionic case is of interest due to the antisymmetry property,
and moreover  the  associated Fock space is finite whenever
$\mathcal{H}$ is finite. \\
Let $\mathcal{H}$ be a complex $n$-dimensional Hilbert space,
 with canonical basis $\{e_i,\;i=1,\ldots,n \}$ describing single particle
 states. Its dual space is  $\mathcal{H}^*$  with   basis
 $\{e^j,\;j=1,\ldots,n \}$.

The fermionic Fock space is given  in terms of  the Grassmann
algebra based on $\mathcal{H}^*$
\begin{equation}
\mathcal{F}(\mathcal{H^*})={\bigoplus_{n \in \mathbb{N}}S_{-}
(\mathcal{H^*}^{\otimes n})}= \mathbb{C} \oplus \mathcal{H^*} \oplus
\wedge^2 \mathcal{H^*} \oplus... \oplus \wedge^n\mathcal{H^*}
\end{equation}
where $\mathbb{C}$ is the ray of vacuum state, $\mathcal{H^*}$ are
 representing one-particle subspace, $\wedge^2 \mathcal{H^*}$  are representing
 2-particle subspace, etc. The dimensionality is given by
\begin{equation}
dim~ \wedge^k \mathcal{H^*}= \binom{n}{k} \qquad \text{and} \qquad
dim~ \mathcal{F}(\mathcal{H^*})= 2^n.
\end{equation}
Moreover, we can form a $2n$-dimensional complex space
$E=\mathcal{H} \oplus \mathcal{H}^*$ with basis
 $\{e_I\} \equiv \{ e_i,e^j\}$ where $I=\{1,\ldots,n,n+1,\ldots,2n\}$.

Accordingly, one may associate an operator to every element of $E$
to act on $\mathcal{F}(\mathcal{H^*})$, by the following
correspondence
\begin{equation}
e^{i} \mapsto \hat{e}^{i} \equiv e^{i}\wedge, \qquad  \qquad e_j
\mapsto \hat{e}_j \equiv \iota_{e_j}.
\end{equation}
where the mappings are respectively the exterior and interior
products seen as ladder operators on $\mathcal{F}(\mathcal{H^*})$.
Hence, they naturally obey the usual anticommutator relations
\begin{equation}
\{\hat{e}^{i},\hat{e}_j\} = \delta_j^i,  \quad  \quad
\{\hat{e}^{i},\hat{e}^{j}\} = \{\hat{e}_i,\hat{e}_j\} = 0.
\end{equation}
In this way, one may see $\hat{e}^{i}$ and $\hat{e}_j$ as creation
and annihilation operators respectively. Thus,  one may redefine
them as $\hat{e}^{i} \equiv \hat{p}^{i}$ and $\hat{e}_j \equiv
\hat{n}_j$
where $\hat{p}$ is the momentum operator, and $\hat{n}$ is the number operator. \\

Taking $|0\rangle$ as a notation for the vacuum state, we have
$\hat{n}_j|0\rangle=0$   which is the defining property for vacuum.
$\hat{p}^{i}|0\rangle$ ,
   $\hat{p}^i\hat{p}^j|0\rangle$ , $\hat{p}^i\hat{p}^j\hat{p}^k|0\rangle$ ... ,
$\hat{p}^1\hat{p}^2...\hat{p}^n|0\rangle$ with $(i \leq j \leq k)$
 represent respectively a single particle in the $i^\text{th}$ mode,
 and similarly two, three .., $n$-particle states. \\
We note that a $k$-particle subspace is spanned by the basis vectors
$\hat{p}^{i_1}\hat{p}^{i_2}...\hat{p}^{i_k}|0\rangle$ with
 $(1\leq i_1 \leq i_2 \leq ... \leq i_k \leq n)$. Hence for a general state
$\psi \in  \mathcal{F}(\mathcal{H^*})$ we have
\begin{equation}
\psi= \hat{\Psi}|0\rangle \quad \text{where} \quad
\hat{\Psi}=\sum_{k=0}^n{(\sum_{i_1
i_2...i_k=0}^n{\frac{1}{k!}\psi^{(k)}_{i_1
i_2...i_k}\hat{p}^{i_1}\hat{p}^{i_2}...\hat{p}^{i_k}})}
\end{equation}
where   $\psi^{(k)}_{i_1 i_2...i_k}$ is an antisymmetric tensor of
order $k$ that captures the complex amplitudes of the $k$-particle
subspace.

In particular,  the fermionic Fock space can be decomposed as follows
\begin{equation}
{\cal F}={\cal {F}}_ +\oplus {\cal {F}}_ -
\end{equation}
where the subspaces ${\cal {F}}_-$ and  ${\cal {F}}_+ $ are
associated, respectively,  with positive and negative chiralities. This space has
 been investigated in connection with qubit systems \cite{24}.
A close inspection shows that the  fermionic  Fock space   can be
associated with space of forms $H$ structure. The  latter
contains $2^n\times 2^n$ $(p,q)$-forms. These forms can belong to
two classes called odd and even forms according to the parity of the
number  $p+q$. Indeed, we have the following sectors
\begin{eqnarray}
H^+&=&\{(p,q)-\mbox{forms  where  $p+q$ is an even number} \}\\
H^-&=&\{(p,q)-\mbox{forms  where $p+q$ is an odd number} \}.
\end{eqnarray}
This factorization can be also  understood in terms of  a $\mathbb{Z}_2
$ symmetry  acting on the corresponding complex variables $z$ as follows
\begin{equation}
 z_i \to  -z_i, \qquad i=1,\ldots, n.
 \end{equation}
 The calculation shows that $H$  splits into positive and negative
eigenspaces of  the  $\mathbb{Z}_2$  operator
\begin{equation}
H={ {H}}_ +\oplus {{H}}_ -
\end{equation}
 The splitting of  $H$   leads to equal dimensions for positive and negative
eigenspaces.\\
 Inspired by the BHQC,   the total cohomology space $H$ can be identified
 with  the fermionic Fock space
\begin{equation}
 H \equiv {\cal F}.
 \end{equation}
 In this way, an arbitrary state of ${\cal F}$  can be related   to  an  element of $H$.
 In this interplay, the one dimensional  subspace  associated with
 the vacuum state corresponds to the cohomology subspace  $H^{0,0}$. The other
 states can be associated with the cohomology subspaces  $H^{p,q}$
 where $(p,q)\neq(0,0).$   We expect that this   link  could  be explored  to discuss
 qubits in terms of black hole charges  in type II superstrings on
 toroidal
complex  manifolds. This   produces  the following  identifications
\begin{equation}
{\cal {F}}_ -\equiv{ {H}}_ - \qquad {\cal {F}}_  +\equiv  {{H}}_ +.
\end{equation}
These identifications  can be supported by mapping the $(p,q)$-forms
of the $H$ space, carrying information on black hole charges,  to
states, considered as   basis vectors, of the fermionic Fock space. \\
For the cohomology subspaces  $H^{p,q}$ where $(p,q)\neq(0,0)$, the mapping  reads  as
\begin{equation}
(p,q)\mbox{-form} \rightarrow |i_1\ldots,i_p,j_1\ldots,j_q\rangle
\end{equation}
with antisymmetric properties.  Inspired by the  closed string
theory spectrum,  these states can be obtained by operators acting
on the vacuum state associated with the cohomology subspace
$H^{0,0}$. Using  a useful notation
\begin{equation}
i=(i_1\ldots,i_p)\qquad j=(j_1\ldots,j_q),
\end{equation}
these states  can be associated with the following creation
operators
\begin{equation}\label{formstate}
|(i,j)\rangle=\prod_{a=1}^{p}\alpha_a^{i_a}\prod_{b=1}^{q}\tilde{\alpha}_b^{j_b}|0\rangle,
\end{equation}
where $\alpha_0^{0}\tilde{\alpha}_0^{0}$  is  considered as the
identity operator.  $i_k$ is a binary number taking either 1 or 0
according to wether the corresponding variable is present or not.\\
In type II superstrings, the states $|(i,j)\rangle$ are associated with
D$(i+j)$-brane charges where
\begin{equation}
i+j=\sum_{a=1}^{p}{i_a}+\sum_{b=1}^{q}{j_b}.
\end{equation}
The corresponding black hole charges can be obtained from  the gauge
invariant $(i+j+2)$ form field strength of $(i+j+1)$ forms
$D_{i+j+1}$  coupled to D$(i+j)$-branes.  The gauge fields
$A^\alpha$ can be obtained from the decomposition
\begin{equation}
D^\alpha_{i+j+1} \to  A^\alpha \wedge \omega_{ij}
\end{equation}
$w_{ij}$  are  $(i+j)$-forms  on  the associated toroidal
compactification.
 The integration of
the field strength ${dD}_{i+j+1}^{\alpha}$    gives the black hole
charges
\begin{equation}
p^{\alpha }=\int_{S_{\infty }^{2}\times
C_{ij}}\mathcal{F}_{2}^{\alpha }\wedge\omega_{ij}
\end{equation}
where $C_{ij}$  are $(i+j)$-cycles on which  D$(i+j)$-branes  are
wrapped to produce black holes in lower dimensions.
\subsection{ One-qubit }
The 1-qubit system associated with $n=1$ can be represented by
$(p,q)$-forms on $T^2=T_\mathbb{C}^1$. This compactification produces an  $N
= 2$ supergravity in eight dimensions. In the   case of  type IIA
superstring, the relevant objects  are D0 and D2-branes producing
black hole charges in such a compactification. However, the
situation in type IIB is somewhat different. The corresponding black
hole charges should be obtained from D1-branes. Indeed, we obtain two
copies: One of them corresponds to the positive eigenspace
represented by ${ {H}}_ +$ associated with type IIA superstring, while
the second copy corresponds to  negative  eigenspace ${{H}}_-$ associated with
type IIB superstring.\\
In this way, for the first case the D0 and D2-brane charges require that the 1-qubit in
type IIA superstring reads  as
 \begin{equation}
 |\psi_+ \rangle=a_1\alpha_0^0\tilde{\alpha}_0^0|0\rangle+a_2\alpha_1^1\tilde{\alpha}_1^1|0\rangle
 \end{equation}
associated with (0,0) and (1,1) even  forms given by
 \begin{equation} 1,\quad dz_1\wedge
 \overline{dz}_1.
 \end{equation}
 The corresponding black hole charges can be obtained from  the gauge
invariant 2-form  and 4-from  field strengths of  1-form  and 3-from
{\bf R-R} gauge fields  respectively. They are   coupled to D0 and
D2-branes.  The integration of these field strengths    gives the
black hole charges
\begin{equation}
p^1=\int_{S_{\infty }^{2}}\mathcal{F}_{2},\qquad p^2=\int_{S_{\infty
}^{2}\times C_{11}}\mathcal{F}_{2}\wedge\omega_{11}
\end{equation}
where $w_{11}$  is   the  $2$-form   associated with $dz_1\wedge
 \overline{dz}_1$.
 While the second copy of 1-qubit associated with D1-brane
charges in  type IIB superstring reads as
 \begin{equation}
 |\psi_-\rangle=b_1\alpha_1^1|0\rangle+b_2\tilde{\alpha}_1^1|0\rangle
 \end{equation}
 associated with odd  forms on  $T^2=T_\mathbb{C}^1$, namely  with
\begin{equation}
 dz_1,\quad
 \overline{dz}_1.
 \end{equation}
It is worth noting that this copy can be identified with the one
found in \cite{16} associated with the complex structure of the
elliptic curve  $T_\mathbb{C}^1$.  However, the  type IIA copy is
intimately related  to  its K\"{a}hler deformation. We expect that
these copies  could be related by mirror symmetry duality. This
connection deserves to be discussed in  higher dimensional
compactification.

\subsection{ Two-qubit systems}
The 2-qubit case, which is  of relevance  for  entanglement, appears
in the compactification of  type II superstring on $T_\mathbb{C}^2$.  To
identify the corresponding differential forms and D-branes, we use
the Hodge diagram properties. \\
In type IIA superstring,  the  Hodge duality allows to select the two copies of 2-qubit system states
associated with ${ {H}}_ +$. These can be obtained from two
different brane systems. The first copy can be obtained from  black
hole charges associated with a system of
$\{D0,\;D2,\;{\mbox{D4-branes}}\}$.   The corresponding state reads
as
 \begin{equation}
 |\psi^1_+
 \rangle=a_1\alpha_0^0\tilde{\alpha}_0^0|0\rangle+a_2
 \alpha_1^1\tilde{\alpha}_1^1|0\rangle+a_2\alpha^1_3\tilde{\alpha}^1_2|0\rangle+a_4\alpha_1^1\alpha^1_2\tilde{\alpha}_1^1\tilde{\alpha}^1_2|0\rangle
 \end{equation}
which is associated with  the following even forms
\begin{eqnarray}
1~,~ dz_1\wedge \overline{dz}_1~,~dz_2\wedge
\overline{dz}_2~,~dz_1\wedge dz_2\wedge\overline{dz}_1\wedge
\overline{dz}_2.
\end{eqnarray}
It is worth noting that  the  double occupancy embedding of  this
2-qubits   can  be  discussed in terms of the complex geometry of
$T_\mathbb{C}^2$ and the  associated D-branes. In  fact, each state
will be represented by two  boxes associated with 2 factors of
$T_\mathbb{C}^1$

\begin{figure}[!h]
\begin{center}
\begin{tikzpicture}
    \matrix (m) [
        matrixstyle,
        column 2/.style={nodes={squarestyle}},
        column 3/.style={nodes={squarestyle}},
    ]
    {
        $\textbf{D0}:$ \&\color{white}{A} \&\color{white}{A} \& $|00\rangle$ \\
        $\textbf{D2}:$ \& $dz_1 \overline{dz_1}$ \& \color{white}{A} \& $|10\rangle$ \\
                $\textbf{D2}:$ \& \color{white}{A} \& $dz_2 \overline{dz_2}$ \& $|01\rangle$ \\
                $\textbf{D4}:$ \& $dz_1 \overline{dz_1}$ \& $dz_2 \overline{dz_2}$ \& $|11\rangle$\\
    };
\end{tikzpicture}
\end{center}
\caption{Geometric interpretation of double occupancy embedding of
2-qubits. }
\end{figure}

 The second  copy can be obtained from  black hole charges associated with a system having
only  D2-branes. The corresponding   state reads as
 \begin{equation}
 |\psi^2_+
 \rangle=b_1\alpha_1^1\alpha_2^1|0\rangle+b_2
 \alpha_1^1\tilde{\alpha}_2^1|0\rangle+b_3\alpha_2^1\tilde{\alpha}_1^1|0\rangle+b_4\tilde{\alpha}_1^1\tilde{\alpha}^1_2|0\rangle,
 \end{equation}
which is associated with  the following even forms
\begin{eqnarray}
 dz_1\wedge {dz}_2~,~ dz_1\wedge
\overline{dz}_2~,~ dz_2\wedge
\overline{dz}_1~,~\overline{dz}_1\wedge \overline{dz}_2.
\end{eqnarray}
The  negative eigenspace ${ {H}}_ -$  can be  associated with type
IIB superstring in the presence of D1 and D3-branes. In fact, we
obtain two dual  D-brane system  linked by the   complex conjugate
operation:\\
 The first copy is
\begin{equation}
 |\psi^1_- \rangle=c_1\alpha_1^1|0\rangle+c_2
 \alpha_1^2|0\rangle+c_3\alpha_2^1\tilde{\alpha}_1^1\tilde{\alpha}_2^1|0\rangle+c_4\alpha_1^1\tilde{\alpha}_1^1\tilde{\alpha}^1_2|0\rangle
 \end{equation}
and the corresponding odd forms are given by
\begin{eqnarray}
 dz_1~, ~ {dz}_2~,~dz_2\wedge
\overline{dz}_1\wedge \overline{dz}_2~,~dz_1\wedge
\overline{dz}_1\wedge \overline{dz}_2
\end{eqnarray}
The complex conjugate operation  produces the second copy of 2-qubit
\begin{equation}
 |\psi^2_-
 \rangle=d_1\tilde{\alpha}_1^1|0\rangle+d_2\tilde{\alpha}_2^1|0\rangle+d_3\alpha_1^1{\alpha}_2^1\tilde{\alpha}_2^1|0\rangle+d_4
{\alpha}_1^1{\alpha}_1^2\tilde{\alpha}^1_1|0\rangle
 \end{equation}
where  the corresponding forms are
\begin{eqnarray}
\overline{ dz_1}~,~ \overline{{dz}_2}~,~{dz}_1\wedge {dz}_2\wedge
\overline{dz_2}~,~{dz}_1\wedge {dz}_2\wedge\overline{dz_1}.
\end{eqnarray}
The corresponding boxes  (single and mixed occupations)  can be easily
represented by forms on $T_\mathbb{C}^2$.
\subsection{Higher dimensional  qubit systems}
Higher dimensional qubits  can be approached using the same dictionary
of even and odd D-branes of type II superstrings. Accordingly,
three qubit systems have been extensively investigated   in
connection with STU black holes in type IIB superstring. This case
involves more assumptions and calculation based on  the Hodge
duality.\\
In fact, we can identify four  copies of 3-qubit systems
obtained from type IIA superstring  associated with  positive
eigenspace ${ {H}}_ +$. One copy is related to a  type IIA
superstring  black hole obtained from a  system of D0, D2, D4 and
D6-branes. The Hodge diagram shows that  one  could have three
systems of type IIA superstring  black holes obtained only from D2
and D4-branes. In type IIB superstring, there are also four copies
corresponding to negative eigenspace ${ {H}}_ -$. As in type IIA
superstring, one copy can be obtained only from D3-branes. In fact,
this copy can be related  to  the one  involving  D0, D2, D4,
D6-branes using idea of T-duality  and  mirror symmetry.  This
system could recover the STU black hole. The other three involve
system of D1, D3 and D5-branes.

For the compactification on $T_\mathbb{C}^n$,  calculations show that
there are  $2^n$ copies of $n$ qubit system that can be split as
\begin{equation}\label{nn}
2^n=2^{n-1}+2^{n-1}.
\end{equation}
In this way, $2^{n-1}$ copies are  associated with positive
eigenspace ${ {H}}_ +$ in type IIA  geometry with even D-branes. The
other $2^{n-1}$ copies correspond to  negative eigenspace
 ${ {H}}_ -$  associated with type IIB geometry with odd D-branes.

 It is worth nothing that the  integer $n$  could  be fixed by
 string theory compactification. However,  the complex geometry
 could be explored to approach higher dimensional qubits  by introducing
 Hodge combinatorial numbers.

\section{ Qubits from  black holes on  local  Calabi-Yau manifolds}

In this section, we make contact with  local Calabi-Yau
compactifications in type IIA superstring.  Concretely, we consider
 a special  class of   toric manifolds  satisfying the Calabi-Yau
 condition. It is recalled that a toric manifold  can be expressed in the following form, \be
 \mathbf{X}^{n+1} = \frac{\mathbb{C}^{n+r+1}}{{\mathbb{C}^*}^r},
\label{Vpgen} \ee where the $r$ $\mathbb{C}^*$ actions are given by
\be
 {\mathbb{C}^*}^{r}: x_i \to \lambda^{q_i^a} x_i,\quad\ \   i=1,\ldots, n+r+1;
 \quad\ \  a=1,\ldots,r.
\ee In these expressions, the exponents ${q_i^a}$ can be interpreted
as physical  charges and are assumed to be integers \cite{31}.  The
latters have been extensively studied in the context of geometric
engineering  of quantum field theories embedded in string theory and
related models including F-theory. From a physical point of  view, a
clever way to approach such manifolds  is  possible by using a
two-dimensional $\mathcal{N}=2$ supersymmetric linear sigma model
involving  $n+1+r$ chiral superfields $\Phi _{i}$ with charges
$q_{i}^{a},$ $i=1,\ldots, n+1+r;\;a=1,\dots,r$ under $U(1)^{\otimes
r}$ gauge symmetry \cite{31}. The corresponding local geometry  can
be obtained  by solving the D-term potential ($D^{a}=0$) of  such
$\mathcal{N}=2$ linear sigma model. These equations read as
\begin{equation}
\sum\limits_{i=1}^{n+r+1}q_{i}^{a}|\Phi _{i}|^{2}=R^{a},\qquad
a=1,\dots ,r, \label{BDT}
\end{equation}%
where the $R^{a}$'s are FI coupling parameters and where the $\Phi
_{i}$'s are the associated  scalar fields of the chiral superfield
$\Phi _{i}$. The quotient operation by the  $U(1)^{\otimes r}$ gauge
symmetry produces  $n+1$-dimensional toric variety
\begin{equation}
\mathbf{X}^{n+1}=\frac{\mathbb{C}^{n+r+1}\setminus
S}{{\mathbb{C}^{\ast }}^{r}}, \label{Vpgen}
\end{equation}%
where the $r$ copies of $\mathbb{C}^{\ast }$ actions indexed by $\
a=1,\ldots,r$ are given by
\begin{equation}
{\mathbb{C}^{\ast }}^{r}:\Phi _{i}\rightarrow \lambda
^{q_{i}^{a}}\Phi _{i}\
\end{equation}%
and $\lambda $ is  a non zero complex number.  The local
Calabi-Yau condition is ensured by the constraint
\begin{equation}
\sum\limits_{i=1}^{n+1+r}q_{i}^{a}=0.  \label{cy}
\end{equation}
In fact,   there  are many  examples  used in string theory
compactification. An interesting one is the so
called canonical line bundle over projective spaces. \\
To keep the connection with qubit systems of previous sections, we
consider
 the  canonical line bundle over  $n$ complex dimensional  compact toric manifold
 $\mathbf{M}^n$ given by
\begin{equation}
\mathbf{M}^n=\underbrace{{\bf \mathbb{CP}}^1\times {\bf
\mathbb{CP}}^1\times\ldots\times  {\bf\mathbb{CP}}^1}_n
\end{equation}
In this case, each ${q_i^a}$ produces a projective space ${\bf
\mathbb{CP}}^1$. To illustrate this model, we initially consider the
leading example of $n=1$. The  corresponding local Calabi-Yau  $X^2$  is known by
\begin{equation}
{\cal {O}}(-2) \to {\bf \mathbb{CP}}^1
\end{equation}%
 In  sigma model,  this geometry  can be obtained from  a supersymmetric gauge theory with a $U(1)$ gauge symmetry
and three chiral fields $\Phi _{i}$ with charge $(1,-2,1)$. The
D-term constraint (equation of motion of $V$) reads as
\begin{equation}
|\Phi _{1}|^{2}+|\Phi _{2}|^{2}-2|\Phi _{3}|^{2}={Re}(t).
\end{equation}%
This geometry describes the K\"{a}hler deformation of the $A_{1}$
singularity of the ALE spaces
\begin{equation}
uv=z^{2},
\end{equation}%
where $u,v$ and $z$ are the generators of gauge invariants. They are
realized in terms of the scalar fields as follows
\begin{equation}
u=\Phi _{1}^{2}\Phi _{2},\qquad v=\Phi _{3}^{2}\Phi _{2},\qquad
z=\Phi _{1}\Phi _{2}\Phi _{3}.
\end{equation}
The compact part is ${\bf \mathbb{CP}}^1$ which can be obtained by setting
$\Phi _{3}=0$ and identifying
\begin{equation}
|\Phi _{1}|^{2}+|\Phi  _{2}|^{2}={Re}(t).
\end{equation}
The remaining coordinate  can be identified with  the non compact
direction  ensuing the local Calabi-Yau. A  similar analysis is a
priori possible for  higher dimensional cases. In this case, the
corresponding local Calabi-Yau is known by
\begin{equation}
{\cal {O}}(-2,\ldots,-2) \to \underbrace{{\bf \mathbb{CP}}^1\times {\bf
\mathbb{CP}}^1\times\ldots\times  {\bf \mathbb{CP}}}_n
\end{equation}
The compactification of   type II superstrings on such manifolds
gives supergravity models with $ 2^{5-n}$ supercharges interacting
with an abelian gauge symmetry. The corresponding  abelian gauge
theory can  be obtained from type II D-branes wrapping the
appropriate cycles of such local Calabi-Yau manifolds. These
manifolds have been
investigated in connection with black holes in topological string theory \cite{32}.\\
A close inspection  shows  that the Hodge diagram of $\mathbf{V}^n$
can be related  to the one $T_C^n$ up to  $\mathbb{Z}_2^{n}$
orbifold actions. Concretely,   each $\mathbb{Z}_2$ symmetry affects
only one $T^2$ factor. To get the corresponding Hodge numbers  one
can use  the corresponding trivial fibration. Indeed,  if $M$ and
$N$ are complex manifolds, then the trivial fibration  $M \times N$
has Hodge numbers obtained by the following identity \cite{33}
        \begin{equation}
        h^{(p,q)}(M \times N) = \sum_{\substack{u+r=p \\ v+s=q}}h^{(u,v)}(M)h^{(r,s)}(N)
        \label{ident}.
        \end{equation}
In what follows, we will be concerned  only    with  the untwisted
sector. For illustration, we list the Hodge diagrams for $n=1,2,3$
corresponding to ${\bf \mathbb{Z}}_2$, ${\bf \mathbb{Z}}_2\times
{\bf \mathbb{Z}}_2$ and ${\bf \mathbb{Z}}_2\times {\bf
\mathbb{Z}}_2\times {\bf \mathbb{Z}}_2$, respectively, where the
identity \eqref{ident} has been used
\[
\begin{tabular}{|l|l|l|}
\hline
$n=1$ & $%
\begin{tabular}{lllllll}
& & & $h^{0,0}$ & & &  \\
& & $h^{1,0}$ &  & $h^{0,1}$ & & \\
& & & $h^{1,1}$ & & &
\end{tabular}%
$ & $%
\begin{tabular}{lllllll}
&  &  & $1$ &  &  &  \\
&  & $0$ &  & $0$ &  &  \\
&  &  & $1$ &  &  &
\end{tabular}%
$ \\ \hline
$n=2$ & $%
\begin{tabular}{lllllll}
& &  & $h^{0,0}$ &  &  &\\
& & $h^{1,0}$ &  & $h^{0,1}$ & &  \\
& $h^{2,0}$ & &   $h^{1,1}$  & & $h^{0,2} $& \\
& & $h^{2,1}$ &  & $h^{1,2}$ & & \\
& & & $h^{2,2}$ & & &
\end{tabular}%
$ & $%
\begin{tabular}{lllllll}
& &  & $1$ &  &  &\\
& & $0$ &  & $0$ & &  \\
& $0$ & &   $2$  & & $0$& \\
& & $0$ &  & $0$ & & \\
& & & $1$ & & &
\end{tabular}%
$ \\ \hline
$n=3$ & $%
\begin{tabular}{lllllll}
&  &  & $h^{0,0}$ &  &  &  \\
&  & $h^{1,0}$ &  & $h^{0,1}$ &  &  \\
& $h^{2,0}$ &  & $h^{1,1}$ &  & $h^{0,2}$ &  \\
$h^{3,0}$ &  & $h^{2,1}$ &  & $h^{1,2}$ &  & $h^{0,3}$ \\
& $h^{3,1}$ &  & $h^{2,2}$ &  & $h^{1,3}$ &  \\
&  & $h^{3,2}$ &  & $h^{2,3}$ &  &  \\
&  &  & $h^{3,3}$ &  &  &
\end{tabular}%
$ & $%
\begin{tabular}{lllllll}
&  &  & $1$ &  &  &  \\
&  & $0$ &  & $0$ &  &  \\
& $0$ &  & $3$ &  & $0$ &  \\
$0$ &  & $0$ &  & $0$ &  & $0$ \\
& $0$ &  & $3$ &  & $0$ &  \\
&  & $0$ &  & $0$ &  &  \\
&  &  & $1$ &  &  &
\end{tabular}%
$ \\ \hline
\end{tabular}%
\]
It has been shown that  the invariant forms belong to $H^{k,k}_+$
formed by $\prod_{i=0}^{k}dz_i\wedge d\overline{z_i}$. It has been
calculated that the corresponding Hodge numbers are
\begin{equation}
 dim~H_+^{k,k}=h_+^{k,k}=\frac{n!}{k!(n-k)!}.
 \end{equation}
It is clear that one has the following relation
\begin{equation}
dim~H(T^n_C)= \sum_{k=0}^n h_+^{k,k}= 2^{n}
\end{equation}
associated with $n$ qubit systems.    In this way, the relation
(\ref{nn}) can be replaced by
\begin{equation}
2^n= 1+ (2^{n}-1)
\end{equation}
 showing that there  is only one  copy of $n$-qubit systems in type IIA
 superstring. This system is associated with the principal vertical
 line of  the Hodge diagram. The  black hole charges can be  obtained from  a class of even
 D-branes. In this way, the states given in (\ref{formstate})
 reduce to
 \begin{equation}
|(i,i)\rangle=\prod_{a=1}^{p}\alpha_a^{i_a}\tilde{\alpha}_a^{i_a}|0\rangle.
\end{equation}
These states are  associated with double occupancy embedding of the
$n$-qubit Hilbert space in type IIA superstring.  This can be
illustrated in the following figure.

\begin{figure}[!h]
\begin{center}
\begin{tikzpicture}
    \matrix (m) [
        matrixstyle,
        column 2/.style={nodes={squarestyle}},
        column 3/.style={nodes={squarestyle}},
                column 4/.style={nodes={squarestyle}},
                column 6/.style={nodes={squarestyle}},
    ]
    {
        $\textbf{D0}:$ \&\color{white}{A} \&\color{white}{A} \&\color{white}{A} \& \ldots \&\color{white}{A} \& $|000 \ldots 0 \rangle$ \\
        $\textbf{D2}:$ \& $dz_1 \overline{dz_1}$ \&\color{white}{A} \&\color{white}{A} \& \ldots \&\color{white}{A} \& $|100\ldots0\rangle$ \\
              $\textbf{D2}:$ \& \color{white}{A} \& $dz_2 \overline{dz_2}$ \&\color{white}{A} \& \ldots \&\color{white}{A} \& $|010\ldots0\rangle$\\
                \vdots \&\vdots \&\vdots \&\vdots \&\vdots \&\vdots \&\vdots \\
              $\textbf{D4}:$ \& $dz_1 \overline{dz_1}$ \& $dz_2 \overline{dz_2}$ \& \color{white}{A} \& \ldots \& \color{white}{A} \&$|110\ldots0\rangle$ \\
                \vdots \&\vdots \&\vdots \&\vdots \&\vdots \&\vdots \&\vdots \\
                $\textbf{D2n}:$ \& $dz_1 \overline{dz_1}$ \& $dz_2 \overline{dz_2}$ \& $dz_3 \overline{dz_3}$ \& \ldots \& $dz_n \overline{dz_n}$ \&$|111\ldots1\rangle$ \\
    };
\end{tikzpicture}
\end{center}
\caption{Geometric interpretation of double occupancy embedding of
the $n$-qubit Hilbert space in type IIA superstring, where $z_i$
parameterizes the associated ${\mathbb{CP}}^1$.}
\end{figure}
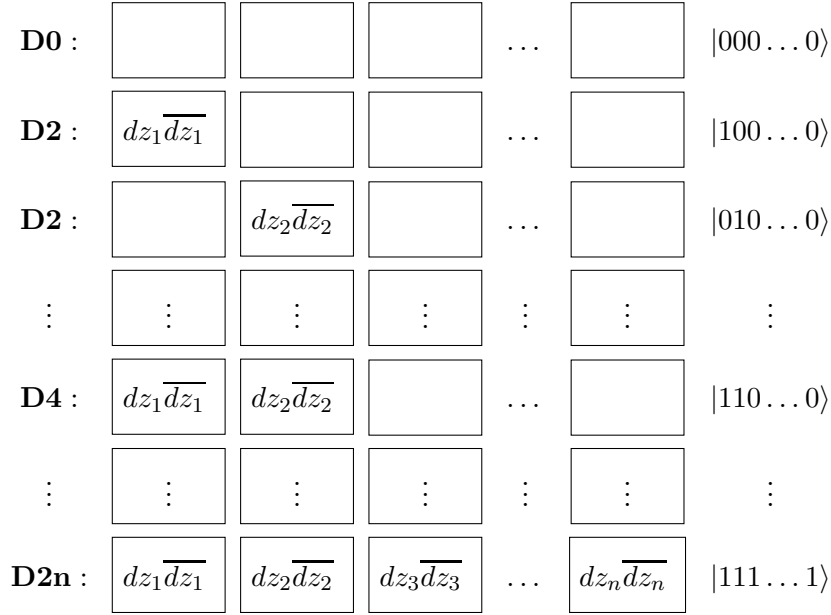

\section{ Conclusion and discussions}

In this paper, we have  investigated a   link between   black holes,
quantum information and fermionic Fock space  using combinatorial
data of  the internal space Hodge diagram.  Concretely, we have
elaborated a one-to-one correspondence between qubit systems, Fock
space  and extremal black holes embedded in maximally supergravity
obtained from  II superstrings compactified on  complex manifolds.
The physical states of $n$-qubit systems can be associated with
differential forms of the  internal  manifolds.  In the
$n$-dimensional toroidal   compactification,  we have shown that
there are  $ 2^{n-1}$ copies of $n$-qubits associated with even
D-branes charges in type IIA superstring. Similar copies can  be
present in type IIB superstring  using odd D-brane charges. Then, we
proposed a possible generalization to Calabi-Yau manifolds. More
precisely, we have shown that  an $n$-qubit system can be associated
with  a canonical line bundle of  $ n$ factors of one dimensional
projective space $ \bf \mathbb{CP}^1$. This qubit system is
associated with type IIA D-brane charges.
 Our paper comes up with many open
questions related to quantum information  geometry. In fact, it
should be of relevance to study   quantum information  using
complex geometry.  It should be interesting to make contact with
mirror symmetry in toric complex manifolds. This matter will be addressed
elsewhere.

{\bf Acknowledgments}: AS is supported by FPA2012-35453.


\begin{thebibliography}{99}
\bibitem{1}
A. Strominger, C. Vafa, {\em  Microscopic Origin of the
Bekenstein-Hawking Entropy}, Phys.Lett.  {\bf B379} (1996) 99, {\tt
arXiv:hep-th/9601029}.
\bibitem{2}
C. Vafa,  {\em Black Holes and Calabi-Yau Threefolds},
Adv.Theor.Math.Phys. {\bf 2} (1998) 207, {\tt hep-th/9711067}.

\bibitem{3} J. Maldacena, A. Strominger, E. Witten,{\ em  Black Hole Entropy in
M-Theory}, JHEP {\bf 9712} (1997)002, {\tt arXiv:hep-th/9711053}.
\bibitem{4}
B. Haghighat, S. Murthy, C. Vafa, S. Vandoren, {\em F-Theory,
Spinning Black Holes and Multistring Branches}, (2015), {\tt
arXiv:1509.00455}.







\bibitem{5}
 S. Ferrara, R. Kallosh, A. Strominger, {\em N = 2 Extremal Black Holes}, Phys. Rev. {\bf
 D52}
(1995) 5412, {\tt hep-th/9508072}.

\bibitem{6}
S. Ferrara and R. Kallosh, {\em Supersymmetry and Attractors}, Phys.
Rev. {\bf D54} (1996) 1514, {\tt hep-th/9602136}.

\bibitem{7}
R. Ahl Laamara, M. Asorey, A. Belhaj, A, Segui, {\em Extremal Black
Brane Attractors on The Elliptic Curve}, J.Phys. {\bf A43} (2010)
105401, {\tt arXiv:0907.0093}.

\bibitem{8}
P. Bueno, R. Davies, C. S. Shahbazi, {\em Quantum black holes in
Type-IIA String Theory}, {\tt arXiv:1210.2817}.

\bibitem{9}
H. Ooguri, A. Strominger, C. Vafa, {\em Black Hole Attractors and
the Topological String}, Phys.Rev.{\bf D70}(2004)106007, {\tt
arXiv:hep-th/0405146}.

\bibitem{10}
S. Bellucci, S. Ferrara, A. Marrani and A. Yeranyan, {\em Mirror
Fermat Calabi-Yau threefolds and Landau-Ginzburg Black Hole
Attractors}, Riv. Nuov o Cim. {\bf029}, 1 (2006), {\tt
hep-th/0608091}.

\bibitem{11} A. Belhaj, {\em
On Black Objects in Type IIA Superstring Theory on Calabi-Yau
Manifolds}, African Journal Of Math. Phys. Vol. 6 (2008)49-54, {\tt
arXiv:0809.1114 [hep-th]}.


\bibitem{12}M. J. Duff, String triality, {\em black hole entropy and Cayley s
hyperdeterminant}, Phys. Rev. {\bf D76} (2007) 025017, {\tt
hep-th/0601134}.
\bibitem{13}
Borsten, M. J. Duff, P. Leevay, {\em The black-hole/qubit
correspondence: an up-to-date review}, {\tt arXiv:1206.3166}.
\bibitem{14} L.
Borsten, M.J. Duff, A. Marrani, W. Rubens, {\em On the
Black-Hole/Qubit Correspondence}, Eur.Phys.J.Plus {\bf 126} (2011)
37, {\tt arXiv:1101.3559}.

\bibitem{15} M. J. Duff, S. Ferrara, {\em Four curious supergravities}, Phys.Rev.
{\bf D83} (2011) 046007, {\tt arXiv:1010.3173}.
 \bibitem{16} P. Levay, {\em  Qubits from
extra dimensions}, Phys. Rev. {\bf D84} (2001) 125020.
\bibitem{17} L. Borsten, D.
Dahanayake, M.J. Duff, W. Rubens, H. Ebrahim, {\em  Freudenthal
triple clas sification of three-qubit entanglement}, Phys. Rev. {\bf
A80} (200 9) 032326,  {\tt arXiv:0812.3322 [quant-ph]}.

\bibitem{18} L. Borsten, D. Dahanayake, M. J. Duff, A. Marrani, W.
Rubens, {\em  Four-qubit entanglement classification from string
theory}, Phys. Rev. Lett. {\bf 105}(2010) 100507, {\tt
arXiv:1005.4915 [hep-th]}.



\bibitem{19}
A. Belhaj, M. B. Sedra, A. Segui , {\em Graph Theory and Qubit
Information Systems of Extremal Black Branes}  J.Phys. {\bf A48}
(2015) 045401, {\tt arXiv:1406.2578}

\bibitem{20}
M. Cvetic, G.W. Gibbons, C.N. Pope {\em Compactifications of
Deformed Conifolds, Branes and the Geometry of Qubits}, (2015), {\tt
arXiv:1507.07585}.
\bibitem{21} Y. Aadel, A. Belhaj, Z. Benslimane, M. B. Sedra, A. Segui, {\em Qubits from Adinkra Graph Theory via Colored Toric
Geometry}, {\tt arXiv:1506.02523}

\bibitem{22}

    A. Belhaj, Z. Benslimane, M. B. Sedra, A.
    Segui, {\em Qubits from Black Holes in M-theory on K3 Surface},   {\tt
    arXiv:1601.07610}.


\bibitem{23} L. Borsten, D. Dahanayake, M.J. Duff, W. Rubens, {\em Superqubits},
Phys. Rev. {\bf D81} (2010) 105023, {\tt
arXiv:0908.0706[quant-ph]}.
\bibitem{24}
P. Levay, F. Holweck, {\em Embedding qubits into fermionic Fock
space, peculiarities of the four-qubit case}, (2015), {\tt
arXiv:1502.04537}.

\bibitem{25} C. Vafa, {\em
Lectures on Strings and Dualities},  {\tt arXiv:hep-th/970220}.

\bibitem{26}
M. A. Nielsen and I. L. Chuang, {\em Quantum Computation and Quantum
Information}, Cambridge University Press, New York, NY, USA, 2000.

\bibitem{27}
D. R. Terno, {\em Introduction to relativistic quantum information},
{\tt arXiv:quant-ph/0508049}.

\bibitem{28}
M. Kargarian, {\em Entanglement properties of topological color
codes}, Phys. Rev. {\bf A78} (2008)062312, {\tt arXiv:0809.4276}.

\bibitem{29} S.W.   Hawking,  {\em Black hole explosions}, Nature {\bf
248}
(1974)(5443)30.




\bibitem{30}
 G. Ottavian, {\em Introduction to the Hyperdeterminant and to the Rank of Multidimensional Matrices}, (2013), {\tt
 arXiv:1301.0472}.

\bibitem{31}  E. Witten, {\em Phases of
N = 2 theories in two dimensions}, Nucl. Phys. {\bf B403} (1993)
159, {\tt hep-th/9301042}.


\bibitem{32} M. Aganagic, D. Jafferis, N. Saulina, {\em
Branes, Black Holes and Topological Strings on Toric Calabi-Yau
Manifolds}, JHEP {\bf 0612} (2006) 018, {\tt arXiv:hep-th/0512245}.







\bibitem{33}
P. Griffiths, J. Harris, {\em Principles of Algebraic Geometry} by
Griffiths and Harris, (1994).

















\end{thebibliography}
\end{document}